\documentclass{ws-ijgmmp}

\usepackage[english]{babel}
\usepackage{amssymb,amsmath,epsfig}
\usepackage{amsmath,graphicx}
\usepackage{palatino}
\usepackage{changes}
\begin{document}

\markboth{Authors' Names}
{Instructions for Typing Manuscripts (Paper's Title)}

%%%%%%%%%%%%%%%%%%%%% Publisher's Area please ignore %%%%%%%%%%%%%%%
%
\catchline{}{}{}{}{}
%
%%%%%%%%%%%%%%%%%%%%%%%%%%%%%%%%%%%%%%%%%%%%%%%%%%%%%%%%%%%%%%%%%%%%

\title{Weak deflection angle by asymptotically flat black holes in Horndeski theory using Gauss-Bonnet theorem}

\author{WAJIHA JAVED}

\address{Division of Science and Technology, University of Education, \\Township, Lahore-54590, Pakistan.\\
\email{wajiha.javed@ue.edu.pk} }

\author{JAMEELA ABBAS}

\address{Department of Mathematics, University of Education, \\ Township, Lahore-54590, Pakistan\\
jameelaabbas30@gmail.com }

\author{YASHMITHA KUMARAN}

\address{Physics Department, Eastern Mediterranean
University, \\ Famagusta, 99628 North Cyprus via Mersin 10, Turkey.\\
y.kumaran13@gmail.com }

\author{AL\.{I} {\"O}VG{\"U}N}

\address{Physics Department, Eastern Mediterranean
University, \\ Famagusta, 99628 North Cyprus via Mersin 10, Turkey.\\
ali.ovgun@emu.edu.tr }

\maketitle

\begin{history}
\received{(Day Month Year)}
\revised{(Day Month Year)}
\end{history}

\begin{abstract}
 The principal objective of this project is to investigate the gravitational lensing
by asymptotically flat black holes in the framework of Horndeski theory
in weak field limits. To achieve this objective, we utilize the Gauss-Bonnet theorem
to the optical geometry of asymptotically flat black holes and applying the
Gibbons-Werner technique to achieve the deflection angle of photons in weak
field limits. Subsequently, we manifest the influence of plasma medium
on deflection of photons by asymptotically flat black holes in the context
of Horndeski theory. We also examine the graphical impact of deflection angle
on asymptotically flat black holes in the background of  Horndeski theory in plasma
as well as non-plasma medium.
   
  \end{abstract}

%\pacs{04.40.-b, 95.30.Sf, 98.62.Sb}


\maketitle

\section{Introduction}
The anecdote of a falling apple fostering Newtonian gravity has been imparted on generations since the year 1666. For centuries from then, physicists have believed that the gravitational force is related to the ratio of the product of the interacting masses and the square of their separation through a proportionality constant. This 'Gravitational' constant was given an accurate value via the Cavendish experiment. In 1910s, Einstein's general theory of relativity transpired suggesting a finite, spherical universe in which the Gravitational constant, $G$ was discerned to exhibit a dependence on mass distribution and size of the universe, so as to account for the inertial forces \cite{Bartelmann:1999yn}. \par
A decade later, Hubble's observations confirmed the Big Bang Theory, essentially establishing that the universe was expanding, and potentially infinite. The extent of the universe that can be observed is limited to a maximum distance determined by the time that light takes to reach Earth from the observable edge. In other words, mass distribution and the size of the universe changes with time, and hence, so does the gravitational constant, $G$ \cite{Clifton:2011jh,Ade:2015ava}. This discrepancy led to the speculation that the effect of $G$ could rather be a scalar field, than a constant number. According to Einstein's formulation, the so-called metric field contains the influence of gravity, and mathematically known as a tensor. Therefore, the idea of consolidating a scalar field due to mass distribution with the metric is called as the scalar-tensor theory  \cite{Capozziello:2011et,Wagoner:1970vr,Blas:2009qj}. \par
One such generic gravitational scalar-tensor theory is the Horndeski theory \cite{Horndeski:1974wa}. Defined for a four-dimensional spacetime, the scalar field is incorporated as a new degree of freedom to formulate the Lagrangian of the system, begetting second-order field equations of motion. This notion is inspired from the Lovelock theory of gravity: relaxing Lovelock's assumptions not only exercises the scalar-tensor theory, but also extends Einstein's theory of gravity. \par
The kinetic term of the Lagrangian characterizes the quadratic derivative of the field and depends on the Lovelock tensor -- which is proportional to the Einstein tensor, $G_{\mu\nu}$ -- encompassing non-minimal coupling between the scalar field, $\varphi$ and curvature \cite{Herdeiro:2015waa}. Hence, the action principle for a 4-D spacetime ($n=4$) becomes \cite{Anabalon:2013oea,Babichev:2017guv}:
\begin{equation}
    \label{ieq}
    I[g_{\mu\nu}, \varphi] = \int \sqrt{-g} \: d^4 x \left[k \left(R-2\Lambda \right) - \frac{1}{2} \left(\alpha g_{\mu\nu} - \eta G_{\mu\nu} \right) \nabla^\mu \varphi \nabla^\nu \varphi \right],
\end{equation}
where, $k \equiv 1/16\pi G$. Here, the first term accounts for the scalar field with non-minimal coupling for matter owing to the Ricci scalar, $R$, and the second term accounts for the Einstein-Hilbert action for gravity owing to the cosmological constant, $\Lambda$. Values of the parameters $\alpha$ and $\eta$ are governed by the positive energy density of matter field.

The above equation is the reduced form of the action analyzed with standard matter, matter fields and non-standard scalar field \cite{Babichev:2017guv}: implementing assumptions - the geometry is static, spherically symmetric and homogeneous with the scalar and metric fields obeying this symmetry in an asymptotically flat space-time - renders this equation as the limiting case of Horndeski theory. Eq. \ref{ieq} is the foundation of the ensuing work. 

When the gravitational force of a massive object bends the trajectory of light that originated from a distant object behind it, gravitational lensing occurs. This is a consequence of general relativity used to understand the universe, galaxies, dark energy and dark matter \cite{Bartelmann:1999yn}. Various authors studied the gravitational lensing by black holes, wormholes, cosmic strings and other objects since the first gravitational lensing observation by Eddington \cite{Badia:2017art,Latimer:2013rja,Elghozi:2016wzb,Ahmedov:2019dja,Turimov:2018ttf,Abdujabbarov:2017pfw,Schee:2017hof,Ghaffarnejad:2016dlw,  Aazami:2011tw,Virbhadra:2007kw,Keeton:2006sa,Keeton:2005jd,Bhadra:2003zs,Cao:2018lrd,Lim:2016lqv,Sultana:2013fda,Fleury:2017owg,Whisker:2004gq,Chen:2009eu,Eiroa:2002mk,Wang:2019cuf,Mao:1991nt,Bozza:2002zj,Sharif:2015qfa,Virbhadra:2002ju,Kasikci:2018mtg,Zhang:2018yzr}.

In 2008, Gibbons and Werner cleverly showed that there is an alternative way to obtain the weak deflection angle for asymptotically flat optical spacetimes using the Gauss-bonnet theorem \cite{Gibbons:2008rj}. Later, Werner managed to derive the weak deflection angle of stationary spacetimes using the same \cite{Werner:2012rc}. Note that both considered the source and the observer to be placed at asymptotic regions. Next, Ishihara et al. showed that it is also possible to use this method for finite-distances (large impact parameter cases) \cite{Ishihara:2016vdc}. Then, Crisnejo and Gallo showed that the plasma medium deflects photons \cite{Crisnejo:2018uyn}. For more recent works, one can see \cite{Ishihara:2016sfv,Arakida:2017hrm,J1 20,J1 21,J1 22,J1 39,Ovgun:2019wej,Ono:2018ybw,Ono:2017pie,Jusufi:2017vta,Ovgun:2018xys,Jusufi:2017vew,Jusufi:2017lsl,Jusufi:2017hed,Ono:2018jrv,Jusufi:2018kmk,Li:2019mqw,Li:2019vhp,deLeon:2019qnp,plasma,Crisnejo:2019xtp,Ovgun:2018prw,Jusufi:2017uhh,Jusufi:2018jof,Ovgun:2018fnk,Ovgun:2018ran,Ovgun:2018oxk,Ovgun:2018fte,Ovgun:2018tua,Javed:2019a,Javed:2019b,Javed:2019jag,Javed:2019rrg,Kumaran:2019qqp,Javed:2020frq,Ovgun:2020gjz,Li:2020wvn,Javed:2020fli,Li:2020dln,Javed:2020lsg,Ovgun:2019qzc}.

Multi-messenger astronomy constrained the scalar-tensor theories substantially through the detection of GW170817. The arrival times of the gravitational waves and it's electromagnetic counterpart from the NGC 4993 galaxy were observed to have fluctuated by less than a minute when two neutron stars spiralling each other ultimately merged. The speed of the gravitational wave is seen to be affected when the scalar field is coupled to curvature. To be coherent with these observations, quintic and quartic models are neglected restricting our calculations to linear observables. Note that Horndeski theories have a serious flaw, related to their primordial tensor spectrum, namely, the gravitational wave speed is not equal to unity. Theories of this sort are problematic. For example, the detection of GW170817 eliminates any late-universe application of Horndeski theory \cite{Ezquiaga:2017ekz}. These theories can be amended by using a new framework developed in \cite{Oikonomou:2020sij}, firstly developed in \cite{Odintsov:2020sqy} and improved by \cite{Oikonomou:2020oil,Odintsov:2020zkl,Odintsov:2020xji}. The results inferred by these studies are however beyond the scope of this work.

In this paper, we intend to study the deflection angle of the black holes governed by Horndeski theory using the Gauss-Bonnet theorem to test the validity of the modified gravity theory. To compare, we consider the idea of the deflection angle of massive particles in a plasma medium from a black hole. Our main aim is to check the effects of Horndeski theory on weak deflection angle.

This paper is organized as follows: section \textbf{$\textrm{2}$} reviews some basics on asymptotically flat black holes, computes the Gaussian optical curvature and calculates the deflection angle using GBT. In section \textbf{$\textrm{3}$}, we calculate the
deflection angle in plasma medium, followed by concluding remarks in the last section.

\section{Calculation of photon lensing for for Asymptotically flat black holes}
When the action comprises of a cosmological term, a new asymptotically locally flat black hole can be found. Here, the kinetic term (constructed with Einstein tensor) of the scalar field alone is considered to yield the matter term taking $\alpha=0$ in the action, which reduces the latter to:
\begin{equation}
I[g_{\mu\nu},\varphi]=\int\sqrt{-g}d^{4}x[k(R-2\Lambda)+\frac{\eta}{2}G_{\mu\nu}\nabla^{\mu}\varphi \nabla^{\nu}\varphi].
\end{equation}

In \cite{Oikonomou:2020sij}, the authors have obtained an equation applying the slow-roll conditions to incorporate the consequences due to the experimental findings of GW170717, further simplified by \cite{Oikonomou:2020oil}: the action appears to acquire an extra term equal to $\varphi \mathcal{G}$ where $\mathcal{G} = R^2 - 4 R_{\mu\nu} R^{\mu\nu} + R_{\mu\nu\rho\sigma} R^{\mu\nu\rho\sigma}$.

Setting the integration constant that emanated from the first integral of the field equation to zero, the following metric defines a solution of the system for $K\neq0$ \cite{Anabalon:2013oea}:
\begin{equation}
ds^{2}=-H(r)dt^{2}+\frac{15(\Lambda r^{2}-2K)^{2}}{K}\frac{dr^{2}}{H(r)}+r^{2}d\Sigma^{2}_{K,2},\\
\end{equation}
where $d\Sigma^{2}_{K,2}=d\theta^{2}+\sin^{2}\theta d\varphi^{2}$ and
\begin{equation}
H(r)=(60K^{2}-20\Lambda Kr^{2}+3\Lambda^{2}r^{4})-\frac{\mu}{r}.
\end{equation}

If $\Lambda$ disappears, the scalar field vanishes reducing the solution to that of the topological Schwarzschild solution in a flat space-time, representing a black hole only in a spherically symmetrical scenario  \cite{Anabalon:2013oea}. By taking $\Lambda=0$:
\begin{equation}
ds^{2}=-H(r)dt^{2}+\frac{60K}{H(r)}dr^{2}+r^{2}d\Sigma^{2}_{K,2},
\end{equation}
and
\begin{equation}
H(r)=60K^{2}-\frac{\mu}{r}.
\end{equation}
Here, $\mu$ is the integration constant and can be explicate as the black hole mass. Now to acquire the null geodesics ($ds^{2}=0$), the black hole optical
spacetime in equatorial plane $\theta={\pi}/{2}$ is written as:
\begin{equation}
\label{nullgeod}
dt^{2}=\frac{60K}{H(r)^{2}}dr^{2}+\frac{r^{2}}{H(r)}d\varphi^{2},
\end{equation}
along with optical metric $\hat{g}^{opt}_{ab}=\frac{g_{ab}}{(-g_{tt})}$, 
according to the Fermat principle, the geodesics are spatial 
rays of light. The Gaussian optical curvature can be determined by 
allowing use of the description for the two-dimensional 
optical metric given in Eq. \ref{nullgeod}
\begin{equation}
\mathcal{K}=\frac{R_{icciScalar}}{2},
\end{equation}
in which $R$ for optical metric is the Ricciscalar. Following the computation of non-zero Christoffel symbols, we obtain 
the following equation particularly for the Gaussian optical curvature 
of the optical metric
\begin{equation} \label{cur}
\mathcal{K}=-{\frac {K\mu}{{r}^{3}}}+{\frac {{\mu}^{2}}{80\,K{r}^{4}}}+\mathcal{O}(r^{-5}).
\end{equation}

Let us recall the GBT for a two dimensional manifold. In this regard,
we consider a regular domain $D_{R}$ aligned by 2-dimensional surface $S$
with Riemannian metric $\hat{g}_{ij}$, along with its piece-wise smooth boundary
$\partial D_{R}=\gamma_{g}\cup C_{R}$, then GBT permits a connection among
the geometry and topology in terms of the subsequent relation \cite{Gibbons:2008rj}
\begin{equation}
\int\int_{D_{R}}\mathcal{K}dS+\oint_{\partial D_{R}}\hat{k}d\sigma+
\sum_{j}\tilde{\theta}_{j}=2\pi \mathcal{X}(D_{R}),
\end{equation}
where $\mathcal{K}$ is the Gaussian optical curvature, $\tilde{\theta}_j$ is the exterior angle at the $j^{\text{th}}$ vertex and $\sigma$ is the line element along the boundary $D_R$. Let $\gamma$ be a smooth curve in the same domain. Thus, $\dot{\gamma}$ comes to be the unit speed vector \cite{Gibbons:2008rj}. It is well
known that for regular domain the Euler characteristic $\mathcal{X}_{D_{R}}=1$,
while $\hat{k}$ is termed as a geodesic curvature and is defined as 
\begin{equation}
\hat{k}=g^{opt}(\nabla_{\dot{\gamma}}\dot{\gamma}, \ddot{\gamma}),
\end{equation}
having the unit speed condition $g^{opt}(\dot{\gamma}, \dot{\gamma})=1$,
where $\ddot{\gamma}$ is the unit acceleration vector perpendicular to $\dot{\gamma}$.
In the case of $R\rightarrow\infty$, the respective jump angles are taken as $\pi/2$ (in short, the sum of angles corresponding to the observer and the source: $\tilde{\theta}_{O}+\tilde{\theta}_{S}\rightarrow \pi$). Using the fact that the geodesic curvature offers no contribution i.e.
$\hat{k}(\gamma_{\tilde{g}})=0$, we shall pursue a contribution by the virtue of the curve $C_{R}$ computed as:
\begin{equation}
\hat{k}(C_{R})=\mid \nabla_{\dot{C}_{R}}\dot{C}_{R}\mid.
\end{equation}
Let us consider $C_{R}:=r(\varphi)=R= \text{const}$, while $R$ endows the distance
from the coordinate origin. The radial component of the geodesic curvature
states as
\begin{equation}
(\nabla_{\dot{C}_{R}}\dot{C}_{R})^{r}=\dot{C}_{R}^{\varphi}(\partial_{\varphi}
\dot{C}_{R}^{\varphi})+\Gamma^{r}_{\varphi\varphi}(\dot{C}_{R}^{\varphi})^{2}.
\end{equation}

Using the above equation, we note that the first term vanishes, then the second term can be obtained using the unit speed condition. Then, $\hat{k} $ is calculated as: $\lim_{R \to \infty} \hat{k} (C_{R}) =\lim_{R\to \infty}\left\vert \nabla_{\dot{C}_{R}}\dot{C}_{R}\right\vert\to\frac{1}{R}$. We take the large limits of the radial distance, and find: $\lim_{R \to \infty } \mathrm{d}t\to  R \, \mathrm{d}\varphi.$ Hence the deflection angle can be calculated in the form: \cite{Gibbons:2008rj}
\begin{equation} \label{gbt}
\Theta=-\int_{0}^{\pi}\int_{b/\sin\varphi}^{\infty}\mathcal{K}dS,
\end{equation}
where $b$ is the impact parameter, a dimensionless quantity that endorses the straight line approximation in which the light ray is assumed to be expressed as $r = b/\sin \varphi$ at zeroth order in the weak deflection limits \cite{Gibbons:2008rj}. This equation inscribes the global impact on the lensing of particles on
account of the fact that one has to integrate over the
optical domain of integration outside the enclosed mass. Now, by using Eq. \ref{cur} into Eq. \ref{gbt}, we obtain the
weak deflection angle for flat black holes in Horndeski theory to be:
\begin{equation} \label{def}
\Theta=\frac{2K\mu}{b}+\frac{\mu^{2}\pi}{320K b^{2}}+\mathcal{O}(\mu^{3}).
\end{equation}
%The weak deflection angle is increasing with  reduced into Schwarzschild deflection angle if we take $K=1$, upto first order term. It can be seen from the second term in the above equation that the second and higher order terms deviate from the Schwarzschild case.

\subsection{Graphical Analysis}
This segment is dedicated to review the impact of deflection angle $\Theta$ on asymptotically flat black holes graphically and to illustrate the physical eminence of these
graphs to examine the influence of curvature constant $K$ and impact parameter $b$
on the deflection angle by analyzing the stable and unstable state of the black hole.
\subsubsection{Deflection angle versus Impact parameter}
For $\mu=2$, the deflection angle is $\Theta$ plotted against the impact parameter $b$ for different values of the curvature constant $K$ in Figure 1.
\begin{center}
\label{daip}
\epsfig{file=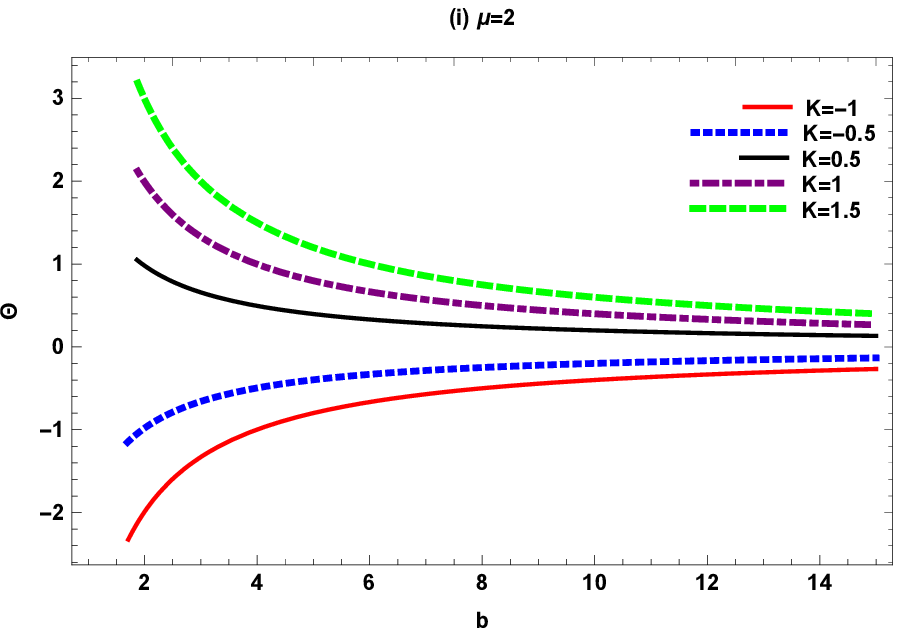,width=0.45\linewidth}\epsfig{file=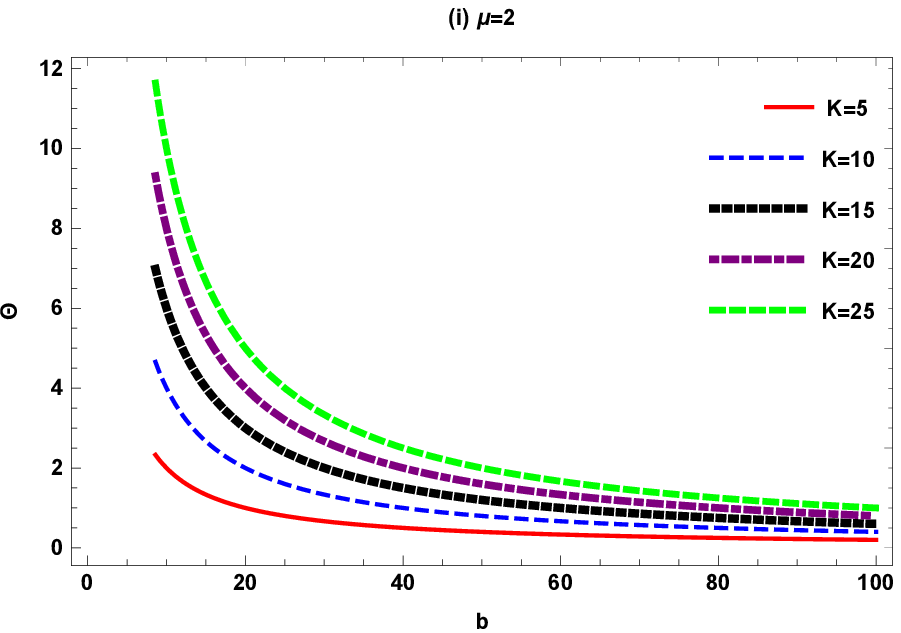,width=0.45\linewidth}\\
{Figure 1: $\Theta$ versus $b$}.\\
\end{center}
\begin{itemize}
\item  \textbf{Figure 1} demonstrates the influence of $\Theta$ w.r.t
$b$ for different values of $K$. One can examine that for small $b$
deflection angle increases but as $b$ increases, the deflection angle decreases for fixed $\mu$.
So for stable behavior we choose the domain $b\in[1,15]$.

Figure (i) illustrates graphically the impact of $\Theta$ w.r.t $b$ by varying $K$. For negative $K$, we obtain locally hyperbolic behavior
but for $K=0$ the behavior is locally flat. If there is small change in the variation
of K, deflection angle is exponentially decreasing.
\item Figure (ii) shows that with the increase of $K$, $\Theta$ decreases gradually as it tends to positive infinity. We obtain physical stable behavior just for $K\geq 0.5$.

\end{itemize}
\section{Photon lensing in a plasma medium}
In this section, we investigate the effect of plasma on photon lensing by asymptotically flat black hole in Horndeski theory.
The refractive index for a flat black hole is stated as follows \cite{Crisnejo:2018uyn},
\begin{equation}
n(r)=\sqrt{1-\frac{\omega_{e}^{2}}{\omega_{\infty}^{2}}H(r)},
\end{equation}
then, the corresponding optical metric illustrated as
\begin{equation}
d\tilde{\sigma}^{2}=g^{opt}_{jk}dx^{j}dx^{k}=\frac{n^{2}(r)}{H(r)}\left[\frac{60K}{H(r)}dr^{2}+r^{2}d\varphi^{2}\right],
\end{equation}
where, the metric function $H(r)$ in the optical metric is given by:
\begin{equation}
H(r)=60K^{2}-\frac{\mu}{r}.
\end{equation}

The corresponding optical Gaussian curvature is calculated by using Eq. $8$ to be:
\begin{eqnarray}
\mathcal{K}&=&-\frac{\mu K}{r^{3}}+\frac{\mu^{2}}{80Kr^{4}}+\frac{90\mu K^{3}}{r^{3}}
\frac{\omega_{e}^{2}}{\omega_{\infty}^{2}}-\frac{9\mu^{2}K}{4r^{4}}\frac{\omega_{e}^{2}}{\omega_{\infty}^{2}}
-\frac{3\mu K}{r^{4}}\nonumber\\&&(60K^{2}r-\mu)\frac{\omega_{e}^{2}}{\omega_{\infty}^{2}}
+\frac{\mu^{3}}{80K r^{5}}\frac{\omega_{e}^{2}}{\omega_{\infty}^{2}}+\frac{3\mu^{2}(60K^{2}r-\mu)}{80K r^{5}}
\frac{\omega_{e}^{2}}{\omega_{\infty}^{2}}.
\end{eqnarray}
Then the geodesic curvature approaches $1$ for $R$ goes to $\infty$ as:
\begin{equation}
\lim_{R\rightarrow\infty}\hat{k}_{g}\frac{d\tilde{\sigma}}{d\varphi}\bigg\vert_{C_{R}}=1.
\end{equation}
Using the straight line approximation given by $r=b/\sin\varphi$ as $R\rightarrow\infty$, GBT can be stated as: \cite{Crisnejo:2018uyn}:
\begin{equation}
\lim_{R\rightarrow\infty}\int^{\pi+\Theta}_{0}\left[\hat{k}_{g}
\frac{d\tilde{\sigma}}{d\varphi}\right]|_{C_{R}}d\varphi=\pi-
\lim_{R\rightarrow\infty}\int^{\pi}_{0}\int^{R}_{b/\sin\varphi}\mathcal{K}dS.
\end{equation}
After simplification, we obtain
\begin{equation} \label{plasmangle}
\Theta\simeq\frac{\mu^{2}\pi}{320b^{2}K}+\frac{\mu^{3}}{90b^{3}K}\frac{\omega_{e}^{2}}
{\omega_{\infty}^{2}}+\frac{2\mu K}{b}-\frac{3\mu^{2}K\pi}{4b^{2}}
\frac{\omega_{e}^{2}}{\omega_{\infty}^{2}}+\frac{180\mu K^{3}}{b}
\frac{\omega_{e}^{2}}{\omega_{\infty}^{2}}\\+\mathcal{O}(\mu^{4},K^{4}).
\end{equation}
The above results shows that the photon rays are moving in a medium of homogeneous plasma.
\subsection{Graphical Analysis}
This section is focused to investigate the graphical effect of deflection
angle $\Theta$ on asymptotically flat black holes in a plasma medium. Further, we exemplify the physical implications of these
graphs to analyze the effect of curvature constant $K$, $\frac{\omega_{e}}{\omega_{\infty}}$ and impact parameter $b$
on deflection angle by analyzing the stable and unstable state of black hole.
\subsubsection{Deflection angle versus Impact parameter $b$}
This subsection offers the examination of deflection angle $\Theta$ w.r.t
impact parameter $b$ for different ranges of curvature constant $K$, $\frac{\omega_{e}}{\omega_{\infty}}$,
for fixed $\mu=2$. For simplicity, here we suppose $\frac{\omega_{e}}{\omega_{\infty}}=\eta$.
\begin{center}
\epsfig{file=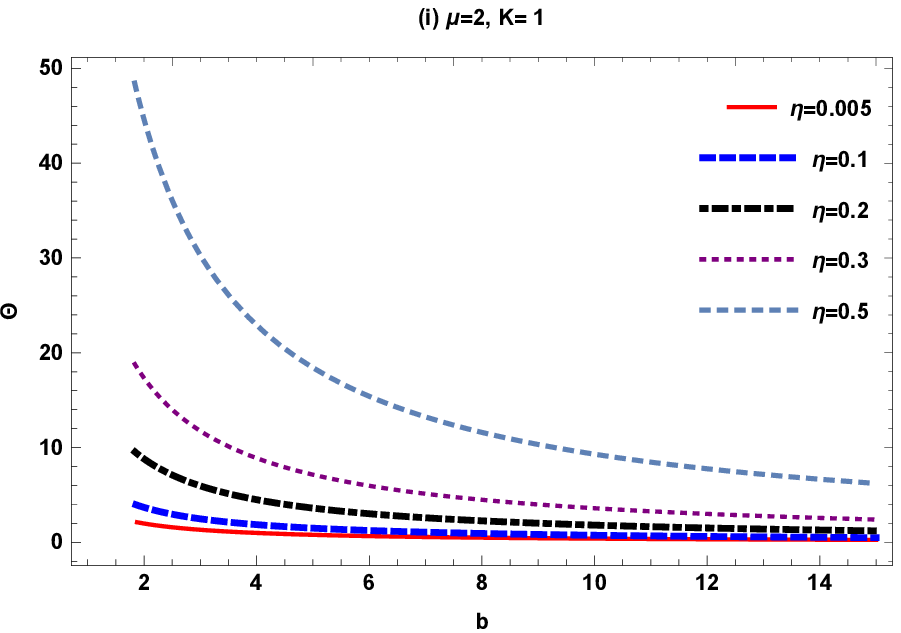,width=0.45\linewidth}\epsfig{file=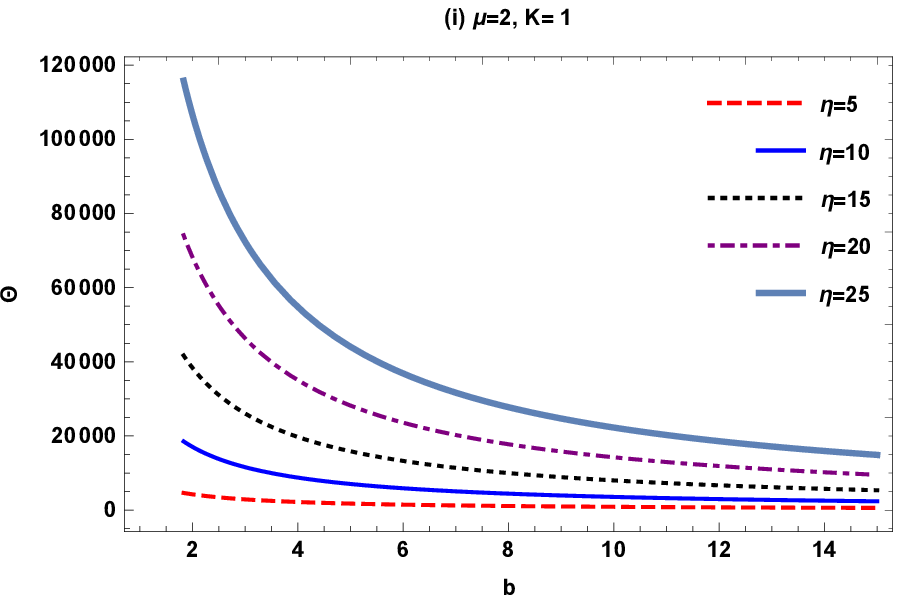,width=0.45\linewidth}\\
{Figure 2: Relation between $\Theta$ and impact parameter $b$}.\\
\end{center}
\begin{itemize}
\item  \textbf{Figure 2} depicts the influence of $\Theta$ w.r.t
$b$ for varying $\eta$ and for fixed $\mu=2$ and $K=1$.
\begin{enumerate}
\item In figure (i), represents the behavior of $\Theta$ w.r.t $b$ for small variation of $\eta$.
For $\eta=0.005$ the deflection angle decreases, it is observed that the behavior is 
same for $\eta=0.005\rightarrow0.01$ and $0.01\rightarrow0.1$ there is small change in 
deflection angle but greater than $0.1$ the deflection angle actually increase.
\item In figure (ii), shows that the deflection angle gradually increase by increasing $\eta$. 
\end{enumerate}
\end{itemize}
\section{Conclusion}
In this paper, we accomplished an extensive analysis of deflection angle of light
by asymptotically flat black hole in the background of Horndeski theory in weak field approximation.
In this regard, we employ the optical geometry of asymptotically
flat black hole in Horndeski theory. Thenceforth, we have utilized the GBT
by using straight line approximation and computed the deflection angle procured by the leading order terms. The obtained deflection angle is evaluated by integrating a domain outside
the impact parameter, that depict the globally impact of gravitational lensing. 
Additionally, we have found the deflection angle of photon lensing for asymptotically flat black hole in a plasma medium. Also, we have analyzed the influence of the impact parameter, the curvature constant and the plasma term on the deflection angle of the photon lensing by asymptotically flat black hole in the context of Horndeski theory graphically. We infer that the proposed deflection angle increases by decreasing the impact parameter, the mass term $\mu$ is found to decrease the deflection angle, and increasing the curvature constant is seen to decrease the deflection angle gradually. 
Moreover, if we disregard the impact of plasma medium $(\frac{\omega_{e}}{\omega_{\infty}}\rightarrow0)$, in the following equation \begin{equation} \label{plasmangle}
\Theta\simeq\frac{\mu^{2}\pi}{320b^{2}K}+\frac{\mu^{3}}{90b^{3}K}\frac{\omega_{e}^{2}}
{\omega_{\infty}^{2}}+\frac{2\mu K}{b}-\frac{3\mu^{2}K\pi}{4b^{2}}
\frac{\omega_{e}^{2}}{\omega_{\infty}^{2}}+\frac{180\mu K^{3}}{b}
\frac{\omega_{e}^{2}}{\omega_{\infty}^{2}}\\+\mathcal{O}(\mu^{4},K^{4}),
\end{equation}
we obtain the weak deflection angle for non-plasma medium case: 
 
\begin{equation} \label{def}
\Theta=\frac{2K\mu}{b}+\frac{\mu^{2}\pi}{320K b^{2}}+\mathcal{O}(\mu^{3}).
\end{equation}

The observations that follow from the Horndeski theory and its mathematical implications include determining observables such as angular positions, separation, magnification and fluxes: a case study of astrophysical applications for Sagittarius A* and M87 can be found in \cite{Badia:2017art}. Additionally, distinct researches include linear Horndeski theories to study dark energy and gravitational waves \cite{Perenon}, and Horndeski gravity to study dark matter \cite{Diez-Tejedor:2018fue}. Our results can be extrapolated to correct for quantum effects while determining the observables, thus, increasing precision - however, the resulting calculations to attain it are beyond the scope of this work. Furthermore, it will be interesting to study the weak gravitational lensing of black holes in MAXWELL– $f(R)$ gravity theories as well as in fourth order gravity theories \cite{Nashed:2019tuk,Horvath:2012kf,Capozziello:2006dp} and non-Minimal Horndeski-like theories (the gravitational wave speed can be equal to unity in natural units, when an appropriate Gauss-Bonnet term is added in the action) using the GBT in future \cite{Oikonomou:2020sij}.

\end{document}